\documentstyle[12pt,aaspp4]{article}

\received{}
\accepted{}
\journalid{}{}
\articleid{}{}

\lefthead{Carpenter & Sanders}
\righthead{The W51 GMC}

\newcommand{\simless}{\mathbin{\lower 3pt\hbox
     {$\rlap{\raise 5pt\hbox{$\char'074$}}\mathchar"7218$}}}
\newcommand{\simgreat}{\mathbin{\lower 3pt\hbox
     {$\rlap{\raise 5pt\hbox{$\char'076$}}\mathchar"7218$}}}

\newcommand{\about}    {$\sim$\thinspace}

\newcommand{\aboutmore}{$\simgreat$\thinspace}

\newcommand{\M}{$^{\rm m}$}
\def\sun{$_\odot$}
\newcommand{\etal}{et~al.\thinspace}
\newcommand{\ts}{\thinspace}

\newcommand{\HH}{H$_{\rm 2}$}
\newcommand{\co}{$^{12}$CO}
\newcommand{\coj}{$^{12}$CO(1--0)}
\newcommand{\thco}{$^{13}$CO}
\newcommand{\thcoj}{$^{13}$CO(1--0)}
\newcommand{\kms}{\thinspace km\kern0.2em s$^{-1}$}
\newcommand{\trstar}{$T_{\rm R}^*$}

\begin{document}

\title{The W51 Giant Molecular Cloud}

\author{John M. Carpenter\altaffilmark{1}\\
email: carp@ifa.hawaii.edu}

\and

\author{D. B. Sanders\\
email: sanders@ifa.hawaii.edu}

\affil{Institute for Astronomy, 2680 Woodlawn Drive, University of Hawaii,
Honolulu, H{\ts}I 96822}

\altaffiltext{1}{Current Address: California Institute of Technology, 
                 Department of Astronomy, MS 105-24, Pasadena, CA 91125; 
                 jmc@astro.caltech.edu}
% Notice that each of these authors has alternate affiliations, which
% are identified by the \altaffilmark after each name.  The actual alternate
% affiliation information is typeset in footnotes at the bottom of the
% first page, and the text itself is specified in \altaffiltext commands.
% There is a separate \altaffiltext for each alternate affiliation
% indicated above.

\begin{abstract}

We present 45\arcsec-47\arcsec\ angular resolution maps at 50$''$ sampling
of the \co\ and \thco\ J=1-0 emission toward a 1.39\arcdeg\ $\times$ 
1.33\arcdeg\ region in the W51 H{\ts}II region complex. These data permit the 
spatial and kinematic separation of several spectral features observed along 
the line of sight to W51, and establish the presence of a massive 
($1.2 \times 10^6~M$\sun), large ($\Delta\ell \times 
\Delta b = 83\ts{\rm pc} \times 114\ts{\rm pc}$) giant molecular cloud (GMC), 
defined as the W51 GMC, centered at ($\ell,b,V)_{\rm c} \sim 
(49.5^\circ,-0.2^\circ,61${\ts}\kms). A second massive ($1.9 \times 
10^5~M$\sun), elongated (136~pc $\times$ 22~pc) molecular cloud is found at 
velocities of \about 68\kms\ along the southern edge of the W51 GMC. Of 
the five radio continuum sources that classically define the W51 
region, the brightest source at $\lambda\ts6\ts$cm (G49.5-0.4) is spatially 
and kinematically coincident with the W51 GMC and three (G48.9-0.3, G49.1-0.4, 
and G49.2-0.4) are associated with the 68\kms\ cloud. Published absorption
line spectra indicate that the fifth prominent continuum source 
(G49.4-0.3) is located behind the W51 molecular cloud. The W51 GMC is 
among the upper 1\ts\% of clouds in the Galactic disk by size and the 
upper 5--10{\ts}\% by mass. While the W51 GMC is larger and more massive than 
any nearby molecular cloud, the average \HH\ column density is not unusual 
given its size and the mean \HH\ volume density is comparable to that in 
nearby clouds. The W51 GMC is also similar to other clouds in that most of 
the molecular mass is contained in a diffuse envelope that is not currently
forming massive stars. We speculate that much of the massive star formation 
activity in this region has resulted from a collision between the 68\kms\ 
cloud and the W51 GMC.

\end{abstract}


\section{Introduction}
\label{intro}

The compact radio continuum sources comprising W51 (\cite{W58}) have long been 
recognized to constitute some of the most luminous star formation regions in 
the disk of the Galaxy (\cite{HL71}; \cite{Hoff71}). The high luminosity,
the large number of inferred O type stars (\cite{Bie75}), and the location of 
these sources within a molecular cloud (\cite{ML79}) all suggest that the W51 
region represents the early formation stages of an OB association. Besides the 
intrinsic interest in the properties of W51, this region represents one of the 
closest analogs in the disk of the Milky Way to the more vigorous star forming 
sites found in other galaxies (e.g. 30~Doradus). Since these latter regions 
are quite distant, W51 affords many advantages in investigating the detailed 
properties of luminous star forming sites and inferring how these regions may 
originate.

One key to understanding the formation and evolution of any star forming 
region is establishing the properties of the molecular cloud out of which the 
stars form. While the molecular gas in the W51 region has been the subject of 
numerous studies, the interpretation of the results remain controversial. 
Scoville \& Solomon~(1973), primarily on the basis of small strip maps in \coj, 
identified several molecular line components toward W51 and derived a 
minimum mass of $10^5$~M\sun\ and a diameter $\sim${\ts}20--30{\ts}pc for the 
molecular cloud that they associated with the most intense radio component at 
$\lambda$\ts6{\ts}cm (G49.5-0.4; \cite{Meh94}). They further suggested that 
this cloud might be physically related to the several thermal radio continuum 
sources that make up the W51 H{\ts}II-region complex (\cite{GS70}; 
\cite{Wil70}; \cite{GG76}). Subsequent studies of the molecular gas toward 
W51 have confirmed the existence of a large molecular cloud (\cite{ML79}; 
\cite{Nak84}; \cite{Ohi84}; \cite{Dame87}; \cite{Sol87}; \cite{Sco87}), 
although various models continue to be proposed for the relationship of the 
multiple spectral features seen in the molecular gas lines and their 
association with the different H{\ts}II regions. 

The primary difficulty in establishing a definitive model of this region
is the unique location of W51 in the Galaxy with respect to the sun. The W51 
region has classically been associated with the tangent point of the Sagittarius
spiral arm, which is oriented such that the line of sight toward W51 
intersects the spiral arm over several kiloparsecs of path length 
(\cite{SB66}; \cite{B70}). Much of the uncertainty surrounding the W51 region 
stems from establishing whether the numerous radio continuum sources and 
molecular clouds represent a single, large massive star forming region, or the 
chance projection of unrelated star forming sites viewed down a spiral arm. To 
better place the W51 region in context with respect to its location in the 
Galactic plane, Figure~\ref{fig1} displays the integrated \coj\ emission in 
10\kms\ velocity bins covering longitudes 40\arcdeg--55\arcdeg\ from the 
Massachusetts-Stony Brook \co\ Survey (\cite{San86}). The W51 region is 
distinguished by bright \co\ emission extending over a 
1\arcdeg$\times$1\arcdeg\ area centered on ($\ell,b$)~\about{\ts}(49.5\arcdeg, 
-0.2\arcdeg) at velocities \aboutmore 55\kms. A ``3--D'' view of the 
($\ell,b,V$) \co\ data cube covering the region surrounding W51 is shown in 
Figure~\ref{fig2}.  The \co\ isointensity contour surface in this figure 
clearly illustrates both the relatively large number of smaller molecular 
clouds with typical internal velocity dispersions of $\Delta V 
\sim${\ts}3--5{\ts}\kms, and the large concentration of \co\ emission 
extending over a \about 20\kms\ interval in the W51 region. Much of the \co\ 
emission in this area has centroid velocities that exceed the maximum velocity 
permitted by pure circular rotation (i.e. $V_{\rm max}$ \about 54--57\kms; 
\cite{BB93}). Such velocities have long been noted in 21~cm H{\ts}I surveys at 
longitudes near $\ell = 50$\arcdeg, and have been attributed to 
large-scale streaming motions of gas in a spiral density wave (e.g. 
\cite{SB66}; \cite{B70}).

In principle the extent and properties of the molecular clouds located in the 
W51 region can be established by using the kinematic information in the 
molecular line data to isolate individual clouds. In practice, previous 
surveys have had either poor resolution or sparse sampling to make such an 
attempt feasible. Therefore, we have obtained full beam sampled maps of the 
W51 region in both \coj\ and \thcoj\ at subarcminute resolution in order to 
determine the relationship between the various molecular components. These 
maps permit us to disentangle the blends of unrelated clouds along the line of 
sight and to obtain more accurate mass estimates of the molecular gas. These 
data can also be compared with similar maps of more nearby clouds that have 
recently been obtained by us and others.

The outline of this paper is as follows. In Section~\ref{obs}, the observing 
procedures are described and channels maps of the \co\ and \thco\ emission are 
presented. Analysis of the different spectral features observed in these maps 
and a more thorough discussion of the main features associated with the 
compact radio continuum sources in W51 is given in Section~\ref{analysis}. In 
Section~\ref{discussion}, we discuss the current massive star formation events 
in the region with respect to the various molecular components and comment on 
the evolution of the W51 Giant Molecular Cloud (GMC). Our conclusions are 
summarized in Section~\ref{summary}.

\section{Observations}
\label{obs}

\subsection{Observing Procedure}

A 1.39\arcdeg~x~1.33\arcdeg\ region (100~x~96~pixels) toward the W51 region
was mapped in \coj\ (115.271203~GHz) and \thcoj \ (110.201370~GHz) in 
April~1995 using the QUARRY receiver array (\cite{Eri92}) on the FCRAO 14~m 
telescope. The full width at half maximum beam size of the 14 meter antenna at 
these frequencies is 45\arcsec\ and 47\arcsec\ at 110 and 115 GHz 
respectively. The data were taken in position switching mode and calibrated 
with the standard chopper wheel method of observing an ambient temperature 
load and sky emission. The backends for each pixel of the array consisted of 
an autocorrelator spectrometer set to span the velocity range from \about 
0--100\kms\ at 78~kHz sampling (0.20\kms\ @ 115~GHz) and 94~kHz resolution 
(0.24\kms). During data reduction the spectra were smoothed to a velocity 
resolution of 0.5\kms.

Previous FCRAO measurements indicate that the spillover and scatter efficiency 
($\eta_{\rm fss}$) of the telescope and radome is \about 0.7 at the observed
frequencies. The observed antenna temperatures corrected by $\eta_{\rm fss}$
are presented as \trstar \ (\cite{KU81}). A further correction, the source 
coupling efficiency ($\eta_{\rm c}$), accounts for the coupling of 
the beam to the source. For a uniform source that fills only the main beam of 
the 14~m telescope, $\eta_{\rm c}$ is \about 0.7 (i.e. 30\% of the power is 
scattered on angular sizes much greater than the FWHM beam size), while for 
sources with uniform intensity over a diameter of 30\arcmin, $\eta_{\rm c}$ is 
1.0. In practice, the observed structures in the \co\ and \thco\ maps span a 
range of sizes and shapes, and applying a single coupling efficiency for the 
entire map is incorrect. For simplicity, we present and analyze the data in 
the \trstar\ temperature scale. The typical rms noise in the \co\ and 
\thco\ maps in 0.5\kms\ channels is $\Delta$\trstar\ \about 0.7~K and 
0.6~K respectively.

\subsection{The Data}
\label{data}

Images of the integrated \co\ and \thco\ intensity ($\int T_{\rm R}^*\ dv)$ 
in 2\kms\ wide intervals are presented in Figures~\ref{fig3} and 
\ref{fig4} respectively in the velocity range from 40 to 70\kms. The 
values printed in the upper left corner of each figure panel denote the 
centroid velocity of the particular interval. Extended \co\ and \thco\ 
emission was detected between 0\kms\ and 25\kms, but these data are not 
presented here. This low velocity emission most likely originates from 
local molecular clouds and is not related to the W51 region of interest here. 
Little emission was observed between 25 and 35\kms\ and at velocities in 
excess of 75\kms\ (see \S\ref{analysis}). Similar velocity structure is also
observed in the 21~cm H{\ts}I emission lines (\cite{B70}). The following section
analyzes the velocity structure in the molecular line maps and identifies
individual molecular clouds.

\section{Analysis}
\label{analysis}

\subsection{Velocity Structure}
\label{velocities}

The \co\ and \thco\ emission toward the W51 region contains a number of 
discrete velocity components that overlap in projection both spatially and 
kinematically. To identify and isolate the emission from these velocity 
components, multiple gaussians convolved with the spectrometer channel widths 
were fitted to each spectrum in an automated manner. The free parameters for 
each gaussian were the amplitude of the spectral line, the mean velocity, and 
the line width. The number of gaussians fitted to each spectrum was determined 
by searching for contiguous channels that contain an integrated intensity 
with a signal to noise ratio $\ge$ 3. Channels containing a local antenna 
temperature maxima (denoted here as channels $c_i, i=1,n$) in each such 
section were then identified. A local maximum at channel $c_i$ was deemed a 
``significant'' peak if the antenna temperature in any channel between c$_i$ 
and the neighboring local maximum at channels $c_{i-1}$ and $c_{i+1}$ 
decreased by more than 2{\ts}$\sigma_{\rm rms}$ from the antenna temperature at 
channel $c_i$. The number of significant peaks corresponded to the number of 
gaussians fitted to that section. Each spectrum was smoothed to a velocity 
resolution of 1.5\kms\ prior to identifying the peaks, although the fits were 
performed on the 0.5\kms\ resolution data. Spectra with large residuals with 
respect to the gaussian fits were visually inspected and additional gaussians 
were added as appropriate. The \thco\ data were easily decomposed into 
gaussians in this manner, but it often became difficult to reliably identify 
the velocity features in the heavily blended \co\ lines. Also, toward the 
compact H{\ts}II regions, some of the structure in the \co\ and \thco\ 
line profiles can be attributed to the absorption of radiation from hot 
molecular gas by colder foreground material (see \cite{ML79}). Away from these 
compact regions, absorption effects are not as significant, and over most of 
the cloud, the peaks in the spectral lines should accurately represent the 
velocity structure along the line of sight.

The results from the gaussian decomposition of the line profiles are 
synthesized in Figure~\ref{fig5}. The upper panel shows histograms of the mean 
velocities for the \co\ (thick lines) and \thco\ (thin lines) gaussians, and 
the lower panel shows the total integrated intensity in the gaussians as a 
function of the mean velocity. Most of the emission is confined to velocity 
intervals of 0--25\kms\ and 35--75\kms. We identify the 0--25\kms\ emission as 
originating from nearby molecular material and the 35--75\kms\ emission with 
molecular gas in the Sagittarius spiral arm. Several velocity components occur 
repeatedly in both the \co\ and \thco\ gaussian fits as signified by the 
histogram peaks shown in the top panel in Figure~\ref{fig5}. In particular, 
velocity components at 7, 15--25, 44, 49, 53, 60, 63, and 68\kms\ are readily 
apparent. In terms of the \co\ and \thco\ integrated intensity, the two major 
features are the 60\kms\ and 63\kms\ components. A 58\kms\ component is 
indicated as well since that is the centroid velocity of the molecular line 
emission toward the brightest radio continuum source in the W51 region 
(\cite{ML79}). Note, however, that the 58\kms\ component is not a prominent 
feature as judged from Figure~\ref{fig5}. The following discussion briefly 
highlights the morphology of the individual velocity components.

\subsubsection{7 and 15--25\kms\ components} 

The 7\kms\ velocity component contains weak, narrow lines over nearly the
entire mapped region and is undoubtedly a nearby molecular cloud. The 
15--25\kms\ interval appears to contain a few distinct velocity features 
(see Fig.~\ref{fig5}), but it is unclear whether or not these components are
physically related.

\subsubsection{44\kms\ component}

The 44\kms\ cloud is elongated parallel to the Galactic plane at $b = 
0$\arcdeg, although this cloud may form part of a larger structure that 
extends to lower Galactic latitudes. The molecular gas at these lower 
latitudes occurs at velocities of \about 40\kms, which is outside the velocity 
range assigned to this feature.

\subsubsection{49 and 53\kms\ components}

The \co\ emission from the 49 and 53\kms\ components is more fragmented than 
the other features mentioned so far. These fragments may represent either 
individual clouds or the remnants of a once larger cloud in the
Galactic plane. The 53\kms\ cloud is distinguished by bright \co\ emission
near $(\ell,b)$ \about $(49.4^\circ,-0.3^\circ)$ that is associated with the 
compact H{\ts}II region G49.4-0.3 (see \S\ref{dis:hii}).

\subsubsection{58, 60, and 63\kms\ components: The W51 Giant Molecular Cloud}

The 63\kms\ component extends for nearly a degree in length and is found mainly 
in the central and eastern part of the mapped region. This is best observed 
in the 66\kms\ panel shown in Figure~\ref{fig3}, which represents the
line wing emission of this velocity component (as well as emission from the
68\kms\ cloud discussed below). The 60\kms\ component consists predominantly 
of a diffuse patch of emission that extends into a filament to the east, and
a second filament to the south. The 60\kms\ and 63\kms\ velocity components, 
along with the 68\kms\ cloud discussed below, likely correspond to the ``high 
velocity stream'' of 21~cm H{\ts}I emission (\cite{B70}) that has been 
attributed to the streaming motions of gas down the Sagittarius spiral arm.

Careful inspection of the channel maps indicates that the spatial distribution 
of the 60\kms\ and 63\kms\ components generally do not overlap. For example, 
the western edge of the 63\kms\ component closely matches the eastern edge of 
the 60\kms\ emission. This is best seen in Figure~\ref{fig4} and the 60\kms\ 
and 66\kms\ velocity panels in Figure~\ref{fig3}. Further, the extended 
emission from the eastern portion of the 63\kms\ component is just above the 
filamentary extension of the 60\kms\ component. Such interfaces are unlikely 
to occur by chance from two unrelated clouds along the line of sight, and 
suggest that the 60 and 63\kms\ components represent kinematic structure 
within a single molecular cloud. Velocity differences of this magnitude are 
commonly observed in nearby molecular clouds (e.g. \cite{Bal87}). 

The 58\kms\ component is dominated by bright, compact molecular emission and 
does not contain the diffuse extended emission that characterizes the 60 and 
63\kms\ features. Inspection of the channel maps suggests that this velocity 
component also reflects the interval kinematic structure within a single cloud 
encompassing the 60\kms\ and 63\kms\ clouds. For 
example, the emission from the filament protruding to the southern portion of 
the mapped region contains primarily a centroid velocity of 58\kms\ closest to 
bright compact \co\ and \thco\ emission. Further away from this bright, 
compact emission region, the velocity of the filament changes to \about 60\kms.
Similar velocity patterns are observed in emission features along the eastern
and western edges of the map. These results suggest that the emission 
constituting the 58, 60, and 63\kms\ components represent the internal 
velocity structure within a single molecular cloud. Koo~(1997) reached similar 
conclusions concerning  the atomic hydrogen clouds at these velocities based 
upon H{\ts}I absorption observations toward the radio continuum sources. Since 
the bright \co\ emission associated with the 58\kms\ and 60\kms\ components 
are coincident with the brightest radio continuum source in the W51 region 
(G49.5-0.5; see \S\ref{dis:hii}), we henceforth refer to the 58--60--63\kms\ 
components as the W51 molecular cloud.

\subsubsection{68\kms\ component}

The 68\kms\ component extends for \about 1\arcdeg\ east--west across the 
image and also likely constitutes part of the ``high velocity stream'' 
identified in H{\ts}I surveys (\cite{B70}). Interestingly, the 68\kms\ 
filament is located at the southern edge of the diffuse emission associated 
with the W51 molecular cloud at velocities \aboutmore 63\kms\ (see 
Figs.~\ref{fig3} and \ref{fig4}). Again, such a clear truncation of 
the W51 molecular cloud at the location of the 68\kms\ cloud is unlikely to 
occur from two random clouds along the line of sight, and strongly suggests 
that these two clouds are physically related objects at a common distance. 
Nonetheless, the elongated appearance of the 68\kms\ cloud is in stark 
contrast to the roughly circular shape of the W51 cloud, indicating that these 
two objects are best treated as individual structures rather than a single 
molecular cloud.

\subsection{Properties of the W51 and the 68\kms\ Molecular Clouds}

The physical properties of the molecular clouds in the W51 region can be 
determined from the gaussian decomposition of the line profiles. We single out
these two clouds since they contain four of the five bright H{\ts}II regions 
found in radio continuum surveys (see \S\ref{dis:hii}). In deriving the 
properties, the distance to the W51 cloud is assumed to be 7.0 $\pm$ 1.5~kpc 
as determined from proper motion studies of the W51{\ts}MAIN ${\rm H_2O}$ 
maser in the G49.5-0.4 dense core (\cite{Gen81}). The 68\kms\ cloud was 
assumed to have the same distance based on its apparent association with the 
W51 cloud as discussed above. The other clouds are not included in this 
analysis since their distances are unknown. 

The properties of the W51 and 68\kms\ clouds are summarized in 
Table~\ref{tbl-1}. The cloud size represents a visual estimate of the 
extent of the detectable \co\ emission along the major ($\theta_{\rm max}$) 
and minor ($\theta_{\rm min}$) axis of the cloud. The cloud line width, 
$\Delta V_{\rm FWHM}$, is the full width at half maximum of the sum of the 
gaussian fits comprising the respective clouds. Two estimates of the cloud 
mass are provided in Table~\ref{tbl-1}. The virial mass, $M_{\rm vir}$, was 
calculated using the expression $M_{\rm vir}~=~209~R~\Delta V^2~M$\sun,
where $\Delta V$ is the full width at half maximum line 
width in kilometers per second and $R$ is the cloud radius ($\equiv 
\sqrt{\theta_{\rm max} \theta_{\rm min}} / 2.0$) in parsecs at the zero 
intensity level. While the cloud size is actually measured at a finite 
\co\ intensity level, we find it unlikely that the clouds are appreciably 
larger at lower intensities, and no correction to the observed cloud size was 
applied. Note that the above expression for the virial 
mass is appropriate for a uniform density, spherical cloud. The equivalent 
expression for a $r^{-1}$ and $r^{-2}$ density cloud would decrease the 
constant factor in the virial mass expression to 188 and 125 respectively.

A second mass estimate can in principal be obtained from the \co\ and \thco\ 
data using the LTE analysis (\cite{Dic78}). However, in blended regions, it 
is often difficult to associate \co\ gaussian fits with analogous \thco\ 
features.  Therefore, the \HH\ column densities were 
estimated by applying a constant conversion factor to the \co\ integrated 
intensities (\cite{Dic75}; \cite{Sol87}; \cite{Strong88}). To ensure that the 
Galactic conversion factor is indeed valid for the W51 region, the conversion 
factor was estimated from lines of sight with unblended \co\ and \thco\ lines 
in the 56--71\kms\ velocity interval that defines the W51 and 68\kms\ clouds. 
The \HH\ column densities for these lines of sight were estimated using the 
procedure outlined by Dickman~(1978) and assuming a \thco/\HH\ abundance of 
1.5~x~10$^{-6}$. A histogram of the ratio of the \HH\ column densities to 
\co\ integrated intensities is strongly peaked with a mean value 
of 1.7~x~10$^{20}$~cm$^{-2}$~(K\kms)$^{-1}$, and is similar to the 
conversion factor that has been derived for the Galaxy (between \about 
2-3~x~10$^{20}$~cm$^{-2}$~(K\kms)$^{-1}$; \cite{Strong88} and 
references therein). The masses ($M_{\rm XCO}$) computed from the \co\ 
integrated intensities were calculated by adopting the conversion factor
derived from the W51 data, and include a multiplicative factor of 1.36 to 
incorporate the mass contribution from heavier elements. 

The results summarized in Table~\ref{tbl-1} indicate that the W51 molecular
cloud has a mean diameter of \about 97~pc and a mass\footnote{The virial mass 
derived here agrees well with 
that estimated by Dame \etal~(1987), but is a factor of 4 less than found by 
Solomon \etal~(1987). Solomon \etal~(1987) derived a substantially larger 
velocity dispersion, and hence computed a larger virial mass, most likely by 
including the emission from the 68\kms\ cloud and several other velocity 
components that we did not associate with the W51 molecular cloud.} 
of \about 10$^6~M$\sun. The 
similarity in the two mass estimates indicate the gravitational potential 
energy is approximately equal to the kinetic energy, and that self--gravity 
must play a critical role in the evolution of the W51 molecular cloud. The 
68\kms\ cloud is 136~pc in length but only 22~pc wide over most of the minor 
axis. Such an obvious departure from spherical symmetry renders the virial 
mass estimate suspect, and so the mass of the 68\kms\ cloud was estimated only 
using the \co\ conversion factor. The mass obtained, \about 10$^5~M$\sun, is 
an order of magnitude less than that of the W51 cloud.

A comparison of the W51 cloud properties with the size and mass spectrum of 
molecular clouds in the Galaxy (e.g. \cite{SSS85}; \cite{Sol87}) shows that 
the W51 molecular cloud is one the largest GMCs in the Galactic disk. Among 
the $\sim${\ts}5000 molecular clouds with diameters in excess of 
$\sim${\ts}22~pc (and corresponding masses $\sim${\ts}10$^5~M$\sun; i.e. 
GMCs), the W51 GMC is in the upper 1{\ts}\% of the clouds by size and the 
upper 5--10{\ts}\% by mass. For the 68\kms\ cloud, although its mass is 
typical of a relatively low mass GMC, a distinguishing feature is its shape. 
The ratio of the major to minor axis of the 68\kms\ cloud is \about 6. In the 
cloud catalog by Solomon \etal~(1987), 85\% of the clouds have an aspect ratio 
less than 2, and only one object has an aspect ratio greater than observed for 
the 68\kms\ cloud. Clearly an elongated shape over such a large length scale 
is unusual in the molecular interstellar material and may indicate that the 
68\kms\ cloud is a transient structure originating from a relatively recent 
dynamical event.

\section{Discussion}
\label{discussion}

\subsection{Massive Star Formation in the W51 Region}
\label{dis:hii}

Now that the individual molecular clouds in the W51 region have been 
identified, their relationship to the massive star forming sites can be 
explored. Three images of the W51 region are shown in Figure~\ref{fig6}:
a map of the integrated \coj\ intensity generated from gaussian fits with mean 
velocities between 56 and 71\kms\ (i.e. the W51 and 68\kms\ clouds), an image 
of the 60\micron\ emission from the IRAS Sky Survey Atlas, and a 
$\lambda$21~cm radio continuum image (\cite{KM97}). The bulk of the 60\micron\ 
emission is elongated parallel to, but slightly below, the galactic plane, and 
is coincident with bright radio continuum emission. Of the sources labeled in 
the $\lambda$21~cm continuum map, W51C has predominantly a non--thermal 
continuum spectrum and is thought to be a supernova remnant (\cite{SG95}), 
while the other sources have thermal spectra and are compact H{\ts}II regions. 
The radio continuum sources G49.4-0.3 and G49.5-0.4 are classically referred 
to as W51A (\cite{KV67}), with the G49.5-0.4 region containing the infrared 
source W51{\ts}IRS1 (\cite{WBN74}) and the ${\rm H_2O}$ masers W51{\ts}North, 
W51{\ts}South, and W51{\ts}MAIN (\cite{Gen77}). Sources G48.9-0.3, G49.1-0.4, 
and G49.2-0.4 are collectively known as W51B (\cite{KV67}). The observed radio 
continuum fluxes (\cite{Koo97}) and far--infrared luminosities (\about 
10$^{6-7}~L$\sun; \cite{Ren84}; \cite{Har86}) imply the presence of one or 
more O stars in each of these regions.

Figure~\ref{fig6} shows that many of the radio continuum sources have 
corresponding peaks in the molecular line and far--infrared images. In 
addition to the spatial coincidences, these radio continuum sources have
recombination line velocities (\cite{Wil70}) similar to the velocity
components identified from the \co\ observations. The G49.5-0.4 H{\ts}II 
region has a recombination line velocity of 59\kms\ and is spatially 
coincident with the strong \co\ emission associated with the W51 molecular 
cloud. The recombination line velocities toward G48.9-0.3, G49.1-0.4, and 
G49.2-0.4 are 66, 72, and 66\kms\ respectively and are located along the 
68\kms\ molecular cloud. Finally, the G49.4-0.3 H{\ts}II region has a 
recombination line velocity of 53\kms\ and is coincident with the bright \co\ 
and \thco\ emission from the 53\kms\ molecular cloud. (Note that the molecular 
gas associated with this H{\ts}II region is outside the velocity range that
defines the W51 GMC.) H{\ts}I (\cite{Koo97} and H$_2$CO (\cite{AG85}) spectra 
toward this source exhibit absorption features at velocities of 63-65\kms,
indicating that G49.4-0.3 must be located behind the W51 GMC. However, these 
observations do not indicate whether this source is situated just beyond W51 
and hence is physically related to the H{\ts}II region complex, or if it is a 
distant, unrelated background massive star forming site. The molecular line 
maps presented here provide no compelling reason for (or against) such an 
association.

To search for any additional massive star forming sites in the W51 molecular 
cloud, the IRAS Point Source catalog was examined for objects that have at 
least two ``high'' quality detections among the four IRAS bands and a 
rising spectral energy distribution toward longer wavelengths. These 
criteria were designed to select a reliable sample of objects with 
far-infrared colors characteristic of embedded star forming regions 
(\cite{Wal89}). Of the 40 IRAS point sources within the mapped region that 
meet these criteria, the seven brightest objects at 25\micron\ are located 
along the interface between the W51 and the 68\kms\ clouds. Visual inspection 
of IRAS 60\micron\ image in Figure~\ref{fig6} confirms that the brightest 
sources are located along this ridge. Assuming that the other point sources 
are located at the distance of the W51 molecular cloud (although many of them 
are almost certainly foreground objects), point sources found away from this 
interface have far-infrared luminosities in the four IRAS bands less than 
40,000~$L$\sun\ and inferred spectral types later than ZAMS B0.5 (\cite{Vac96}).
Thus embedded O type stars in the W51 region currently are confined to the 
southern extreme of the molecular cloud.

\subsection{Comparison to Other Molecular Clouds}
\label{dis:compare}

The extreme star formation characteristics of the W51 region raises the 
question as to whether the star formation activity stems from unusual 
{\it global} properties in the W51 molecular cloud, or from unusual conditions 
found {\it local} to the massive star forming sites. These possibilities 
can be explored by comparing the W51 cloud properties with other 
star forming regions that have also been extensively mapped in \co\ and \thco. 
In particular, we shall compare the W51 and 68\kms\ clouds with the molecular 
clouds associated with the H{\ts}II regions Sh~140, Sh~155, Sh~235, Sh~247, 
Sh~252, and Sh~255 which have been mapped with the same receiver and telescope 
used for the W51 observations (\cite{Heyer96}; \cite{Carp95}). The embedded 
high mass stellar content in this comparison sample is generally limited to a 
single early--B/late O--type star and is in stark contrast to the cluster of 
O stars forming in the W51 molecular cloud.

The clouds in the comparison sample have diameters of \about 20--55~pc and 
masses of \about $2\times10^4$ to $7\times10^4~M$\sun. Thus the W51 molecular 
cloud is \aboutmore 2.5 times larger and \aboutmore 10 times more massive than 
these objects. Note that the molecular clouds associated with Sh~247, Sh~252, 
and Sh~255 are distinct regions within the Gem~OB1 cloud complex, which has a 
total mass of $3\times10^5~M$\sun\ and a diameter of \about 150~pc 
(\cite{Carp95}). While the spatial size of the Gem~OB1 complex is larger than 
the W51 cloud, the W51 cloud exhibits a continuous structure \about 100~pc in 
size as opposed to the fragmentary appearance of the Gem~OB1 complex. The 
68\kms\ cloud is also more massive than the objects in the comparison sample 
and is significantly more elongated than any of the clouds considered here. 
Thus the W51 and 68\kms\ clouds are at the extreme in terms of cloud masses 
and sizes compared to objects in this sample. With the possible exception of 
the 68\kms\ cloud, these clouds are similar though in that they appear to be 
gravitationally bound.

The size and mass of a cloud are not necessarily the key parameters that 
control the star formation activity within the cloud. Intuitively, one might
expect that the amount of matter above a critical density to be the critical 
variable for otherwise similar clouds. Thus if the massive star formation 
activity in W51 has resulted from the large scale collapse of the cloud, one
would expect the volume and column density densities to be larger than found 
in a typical cloud. As an indirect measure of the \HH\ column densities,
Figure~\ref{fig7} shows histograms of the observed \thco\ integrated 
intensities for each of the clouds in our sample. The lowest integrated 
intensity shown for any cloud is 3~K\kms\ since that is approximately the 
highest 3$\sigma$ detection limit among the various \thco\ surveys. Comparison 
of the \co\ and \thco\ intensities suggests that the \thco\ emission is 
optically thin if the \co\ and \thco\ excitation temperatures are equal as 
assumed in the LTE analysis (see \cite{Dic78}). Therefore, the 
distribution of \thco\ integrated intensities should accurately trace the \HH\ 
column density distributions as along as the \thco\ abundance is roughly 
constant within a cloud. The \thco\ integrated intensities in 
Figure~\ref{fig7} can be converted to \HH\ column densities for an assumed 
\thco/\HH\ abundance of 1.5~x~10$^{-6}$ (\cite{BC86}) with the formula 
$${
    N({\rm H_2})~=~3.05\times10^{19}~T_{\rm ex}e^{5.29/T_{\rm ex}}~\int T_{\rm R}^*(^{13}{\rm CO}) 
      dv~~ {\rm cm}^{-2}.
  }$$
For an excitation temperature of $T_{\rm ex}$ = 10~K, the 3~K\kms\ integrated
intensity limit imposed for Figure~\ref{fig7} corresponds to an \HH\ column 
density of 1.5~x~10$^{21}$~cm$^{-2}$, or an visual extinction of \about 1.5\M\ 
(\cite{Boh78}). Variations in the \thco\ abundance and excitation 
conditions will obviously effect the absolute conversion from \thco\ 
integrated intensities to \HH\ column densities, and the comparisons here are 
intended to search for large differences (factors of several or more) in the 
typical column densities in these clouds.

Figure~\ref{fig7} shows that the distributions of \thco\ integrated 
intensities peak near the detection limit of 3~K\kms\ for each cloud, with a 
long tail toward the higher integrated intensities. The tail of these 
distributions correspond to the high column density regions and are often 
associated with star forming sites. The mean \thco\ integrated intensity 
among the clouds varies between 4.9 and 9.7~K\kms, with the W51 cloud 
containing the third highest mean intensity and the 68\kms\ cloud the 
highest. The high values found for the 68\kms\ cloud may be a result of an 
unusual viewing angle, as either this cloud is a sheet of gas observed edge on 
or a long, narrow filament. These results imply that most of the mass in each 
cloud is contained in lines of sight with column densities corresponding to 
less than a few magnitudes of visual extinction. Thus the W51 GMC is similar
to other clouds in that the diffuse envelope contains more mass than the 
high column density cores.

While the W51 cloud contains a higher column density on average than the other
clouds, this can attributed to the fact that it is more than twice as large as
some clouds in the sample. Indeed, assuming that the $1.2\times10^6~M$\sun\ 
W51 cloud is distributed in a sphere of diameter of 97~pc (see 
Table~\ref{tbl-1}), the average \HH\ volume density is 40~cm$^{-3}$, 
comparable to the volume density inferred in nearby molecular clouds 
(\cite{Bli91}; \cite{Carp95}). This suggests that the entire W51 molecular is 
probably not in an advanced stage of collapse, and that the intense star 
formation activity in W51 likely results from forces acting on a localized 
region. In retrospect, this is perhaps not surprising given that the massive 
star forming regions in W51 are located at the edge of the cloud and not in 
the center as expected if, for example, the entire cloud was systematically 
collapsing. 

Contrary to the global properties, the gas properties local to the W51 massive 
star forming regions do appear to be unusual compared to the molecular clouds 
in the solar neighborhood. Submillimeter continuum observations have shown 
that the core containing the G49.5-0.4 H{\ts}II region contains \about 
10$^5~M$\sun\ of gas with a mean \HH\ volume density of \about 
5~x~10$^4$~cm$^{-3}$ over a 3~pc radius region (\cite{Sie91}). By contrast, 
submillimeter observations indicate that the most massive cores in nearby 
molecular clouds typically have masses \about 2--3 orders of magnitude less 
than that of the G49.5-0.4 core (e.g. \cite{Old94}; \cite{Mez92}; 
\cite{Sch87}; \cite{Jaffe84}). While the G49.5-0.4 core does not necessarily 
contain higher gas densities, it does contain more gas at the densities needed 
to form stars.

\subsection{Evolution of the W51 Molecular Cloud}

The above discussion suggests that the key to understanding the massive
star formation activity in the W51 complex is determining the forces 
that acted on a localized region within the cloud. The molecular line maps 
presented here allow us to speculate on what these forces may be. As shown in 
Figures~\ref{fig3} and \ref{fig4}, the diffuse emission from the W51 GMC
truncates at the location of the 68\kms\ cloud for velocities \aboutmore 
63\kms. This morphology was used in \S\ref{analysis} to argue that the W51 and 
68\kms\ clouds are at a common distance since such an interface is unlikely to 
result from the chance superposition of unrelated molecular clouds. One way 
such an interface could form is if the 68\kms\ cloud has collided into the W51 
GMC (see also \cite{AG85}; 
\cite{PPT79}). Molecular gas does extend below the 68\kms\ cloud at lower
velocities, however. In this picture, given the three dimensional structure of 
the W51 GMC, this material has not crossed the path of the 68\kms\ cloud. Both 
\HH CO and H{\ts}I spectra toward the G49.5-0.4 H{\ts}II region (\cite{Koo97}; 
\cite{AG85}; \cite{ML79}) contain absorption lines at velocities \aboutmore 
65\kms, and relative to the line of sight, the 68\kms\ cloud must be located 
in front of the W51 GMC. Since the velocity difference between the two clouds 
(at least 5~\kms) is larger than the sound speed, a shock front will form that 
will compress the molecular gas and possibly induce star formation 
(\cite{EE78}). 

The qualitative model of two colliding clouds accounts for several properties
of the W51 region. First, one would expect that star formation should occur 
preferentially along the interface region. The IRAS image in Figure~\ref{fig6} 
shows that this is indeed the case for massive stars, as the northern half of 
the W51 GMC appears devoid of embedded O type stars despite containing most of 
the cloud mass. The cloud collision model would also suggest that the massive 
star forming regions along the collision interface should have a common age. 
While the ages of these stars are not known, the lifetime of the mid--O type 
stars found in the W51 region (\cite{Meh94}; see also \cite{Koo97}) places an 
upper limit to the stellar ages of \about 5~Myr (\cite{Mey94}). The actual 
ages may be considerably less since these O stars are still in the compact 
H{\ts}II region phase, which has a lifetime of less than 1~Myr (\cite{Chu90}; 
\cite{Com96}). Further support for a cloud--cloud collision model comes from 
considering the projected distance (73~pc) between the two furthest separated 
star forming regions along the W51/68\kms\ cloud interface. The sound travel 
time across this distance is nearly two orders of magnitude larger than the O 
star lifetime. Therefore, these star forming regions must have been created by 
either a single event operating along the entire southern edge of the cloud or 
from up to 4 separate, but nearly simultaneous, events. Given the scarcity of 
embedded O stars in the Galaxy, the cloud--collision model would provide a 
natural explanation for simultaneous star formation along the ridge. 
Ultimately, the suggested collision between the W51 and 68\kms\ clouds may be 
related to a spiral density wave. The anomalous gas velocities in the W51 
region have long been attributed to streaming motions in the Sagittarius 
spiral arm (\cite{SB66}; \cite{B70}). The associated spiral density wave may 
have indirectly led to the massive star formation activity in the W51 region 
by enhancing the number density of clouds and increasing the probability of a 
cloud-cloud collision. Further, a spiral wave shock, if present (see 
\cite{LBC86}), will compress and flatten any clouds (\cite{Woo76}). Indeed, 
such a shock could account for the highly elongated shape of the 68\kms\ 
cloud. The W51 GMC, however, remains roughly circular in shape, and globally 
its evolution is likely still dominated by self gravity. 

Finally, we briefly consider the implications of these results for other star 
forming regions. While the W51 star forming region exceeds all nearby embedded 
star forming sites in terms of the number of O stars, bolometric luminosity, 
and dense core mass, W51 itself is dwarfed by some star forming regions in 
nearby galaxies. Most notably, the 30 
Doradus region in the Large Magellanic cloud contains an order magnitude more 
O stars than W51 (\cite{Vac95}). At larger distances, many interacting 
galaxies contain even yet more vigorous massive star forming regions that may 
be dense enough to represent young globular clusters (\cite{Whi93}). Most 
of the molecular gas appears to have been dispersed in these clusters already, 
and in any event, the distances to these systems precludes any detailed 
studies of the natal clouds. The most significant piece of information 
afforded by the W51 molecular maps is that despite containing one of the most 
massive dense cores known in our Galaxy, most of the mass in the W51 molecular 
cloud is not currently forming massive stars. Thus it is easy to imagine that 
the increase in the number of cloud--cloud collisions that presumably results 
in interacting galaxies may lead to a larger number of star forming regions 
throughout a single cloud or more intense star formation activity within a 
small region. In fact, the mass within the W51 molecular cloud is comparable 
to that in globular clusters, and it may not require the conglomeration of 
many giant molecular clouds to form these stellar systems, but the large scale 
collapse of a single GMC.

\section{Summary}
\label{summary}

We have mapped a 1.39\arcdeg~x~1.33\arcdeg\ region toward the W51 H{\ts}II
region complex at 45\arcsec-47\arcsec\ resolution and 50\arcsec\ sampling in 
the J=1--0 transitions of \co\ and \thco. From these data we have identified 
the major molecular clouds and have associated these clouds with the massive 
embedded star forming sites in the W51 region. We find that:

(1) The two most prominent clouds in the W51 region are the 58-60--63\kms\ 
    cloud (defined as the W51 GMC) and the 68\kms\ cloud. The W51 GMC is 
    associated with the brightest $\lambda${\ts}6{\ts}cm continuum source in 
    W51 (G49.5-0.4), and the 68\kms\ cloud contains the 
    H{\ts}II regions G48.9-0.3, G49.1-0.4, and G49.2-0.4. Published absorption
    line spectra (\cite{Koo97}; \cite{AG85}) indicate that the fifth bright
    H{\ts}II region in the area, G49.4-0.3, must be located behind the W51 
    GMC, but it remains unclear whether or not it is physically associated 
    with the other star forming regions.

(2) The mass of the W51 GMC and the 68\kms\ clouds are \about 
    1.2$\times$10$^6~M$\sun\ and $1.9\times10^5~M$\sun\ respectively. 
    The W51 molecular cloud is roughly circular in shape with a mean 
    diameter of \about 97~pc and appears to be gravitationally bound.  
    Compared to the $\sim${\ts}5000 GMCs in the Galactic disk, W51 is among 
    the top 1{\ts}\% by size and the top 5--10{\ts}\% in terms of cloud mass. 
    The 68\kms\ cloud is an elongated filament of \about 
    136{\ts}pc{\ts}$\times${\ts}22{\ts}pc in size. The 6:1 aspect ratio of the 
    major and minor axis in the 68\kms\ are rare in the Galaxy over 
    such a large size scale (\cite{Sol87}), and suggests that this molecular 
    cloud may represent a transient feature.

(3) The properties of the W51 and 68\kms\ clouds are compared with nearby 
    clouds that have been studied in a similar manner but contain lower
    levels of massive star formation activity. While the W51 cloud is larger 
    and more massive than nearby clouds, the mean \HH\ column density is not 
    unusual given the large size, and the mean \HH\ volume density is 
    comparable. The W51 GMC is similar to other clouds in that most of 
    the molecular mass is contained in a diffuse molecular envelop that is not 
    forming massive stars. The 68\kms\ cloud contains the largest 
    mean column density among the clouds studied here, but this may be a 
    result of an unusual viewing angle for this elongated cloud. We suggest
    that much of the star formation activity in the W51 region has not 
    resulted from global collapse of the W51 cloud, but from forces acting on 
    localized regions within the cloud.

(4) We speculate that much of the massive star formation activity in W51 has 
    resulted from a collision between the W51 and 68\kms\ molecular clouds. 
    This conjecture can explain the string of embedded O stars that are spread 
    out for 70~pc along the interface between the W51--68\kms\ clouds, and why 
    massive star formation is currently confined to the southern ridge of the 
    W51 GMC.
%\vskip 2truein

\acknowledgments

We would like to thank Mark Heyer for completing the \thco\ map of W51 and 
Bon--Chul Koo for making available his $\lambda$21~cm continuum image of W51. 
JMC acknowledges support from the James Clerk Maxwell Telescope Fellowship. 
DBS was supported in part by NASA grant NAGW-3938.  
The Five College Radio Astronomy Observatory is operated with support from
NSF grant 94--20159.

\newpage
\figcaption[Carpenter.fig1.ps]
{   
    \label{fig1}
    Galactic longitude--latitude images of $^{12}$CO(J=1--0) integrated 
    intensity ($\int T_{\rm R}^* dv$) in 10~km{\ts}s$^{-1}$ intervals along 
    the Galactic midplane ($\vert b \vert \leq 1\arcdeg$) from $\ell = 
    40\arcdeg-55\arcdeg$. The data are from the Massachusetts--Stony Brook 
    Galactic Plane Survey (Sanders et al.~1986). The W51 GMC complex is seen 
    most clearly in the $\Delta V = 55-65$ km{\ts}s$^{-1}$ panel as bright 
    $^{12}$CO emission that extends over $\sim$ 1$^\circ$~x~1$^\circ$ centered 
    near $\ell = 49.5\arcdeg, b = -0.2\arcdeg$. A high velocity ``ridge'' of 
    $^{12}$CO emission is evident in the $\Delta V = 65-75$ km{\ts}s$^{-1}$ 
    panel at the same $\ell$ and $b$.
}

\figcaption[Carpenter.fig2.ps]
{   
    \label{fig2}
    A 3-dimensional, velocity--Galactic longitude--Galactic latitude 
    ($v,\ell,b$), isosurface plot of $^{12}$CO(1-0) emission at the 
    $T_{\rm R}^*$ = 4~K level for the region of the Galactic Plane containing 
    W51. The data are from the Massachusetts--Stony Brook Galactic Plane 
    Survey (Sanders et al.~1986).
}

\figcaption[Carpenter.fig3.ps]
{   
    \label{fig3}
    Channel maps of the integrated $^{12}$CO(J=1--0) emission in 
    2~km\kern0.2em s$^{-1}$ intervals over the velocity range from 
    40--70~km\kern0.2em s$^{-1}$. The values printed in the upper right corner 
    of each panel denote the center of the velocity intervals. The grey scales 
    in each panel are displayed in a logarithmic stretch from 
    log$_{10}$(1.0 km\kern0.2em s$^{-1}$) [white] to
    log$_{10}$(10.0 km\kern0.2em s$^{-1}$) [black].
}

\figcaption[Carpenter.fig4.ps]
{   
    \label{fig4}
    Same as in Fig.~3, except for $^{13}$CO(J=1--0). The grey scales in each 
    panel are displayed in a logarithmic stretch from 
    log$_{10}$(0.6 km\kern0.2em s$^{-1}$) [white] to
    log$_{10}$(5.0 km\kern0.2em s$^{-1}$) [black].
}

\figcaption[Carpenter.fig5.ps]
{
    \label{fig5}
    {\it Top:} Histograms of the number of gaussians in the $^{12}$CO and 
    $^{13}$CO data cubes as a function of the fitted mean velocity.
    {\it Bottom:} The total integrated intensity in the gaussian components 
    as a function of the fitted mean velocity. The designated peaks denote a
    velocity component that occurs frequently in the $^{12}$CO and 
    $^{13}$CO images, and the brackets show the range of velocities
    subjectively assigned to each component. The 58~km\kern0.2em s$^{-1}$ 
    velocity component is not a prominent feature as judged from this figure, 
    but is included nonetheless since that is the velocity of the molecular 
    gas associated with the brightest radio continuum emission.
}

\figcaption[Carpenter.fig6.ps]
{   
    \label{fig6}
    Various images of the W51 region.
    {\it Left:}   Map of the $^{12}$CO(1--0) integrated intensity 
                  ($\int T_{\rm R}^* dv$) for gaussian fits with mean velocities 
                  between 56 to 71~km\kern0.2em s$^{-1}$. The contour levels
                  are 10, 25, 50, 100, and 200~K~km\kern0.2em s$^{-1}$.
    {\it Center:} Grey scale image of the 60$\mu$m emission from the
                  IRAS Sky Survey Atlas. The contours begin at 
                  log$_{10}$(100 MJy/ster) with logarithmic intervals of 0.15.
    {\it Right:}  Map of the $\lambda${\ts}21{\ts}cm radio continuum emission 
                  from Koo \& Moon~(1997). The contours levels are 0.015, 0.06, 
                  0.12, 0.3, 1.0, and 2.4~Jy~beam$^{-1}$. The dashed circle 
                  represents the 0.1 primary beam attenuation level, although 
                  the image itself has not been corrected for primary beam
                  response. The prominent radio continuum sources are 
                  labeled in the $\lambda${\ts}21{\ts}cm continuum image. 
                  The H{\ts}II 
                  regions G49.4-0.3 and G49.5-0.4 are classically 
                  referred to as W51A, and G48.9-0.3, G49.1-0.4, and 
                  G49.2-0.4 as W51B. W51C has a nonthermal spectrum and 
                  is likely a supernova remnant.
}

\figcaption[Carpenter.fig7.ps]
{   
    Histogram of the $^{13}$CO(1-0) integrated intensity for W51, the 
    68 km\kern0.2em s$^{-1}$ cloud, and clouds associated with various H{\ts}II 
    regions that have been mapped in the similar manner (Carpenter, Snell, \& 
    Schloerb~1995; Heyer, Carpenter, \& Ladd~1996). The histograms have been 
    truncated at an intensity of 3~K~km\kern0.2em s$^{-1}$ since that is the 
    highest 3$\sigma$ detection limit among the various surveys. Assuming that 
    the $^{13}$CO emission is optically thin, the $^{13}$CO abundance relative 
    to \HH\ is approximately constant, and the excitation conditions are 
    similar, these histograms should reflect the relative \HH\ column density 
    distributions in these clouds.
    \label{fig7}
}

\begin{table}
\dummytable\label{tbl-1}
\end{table}

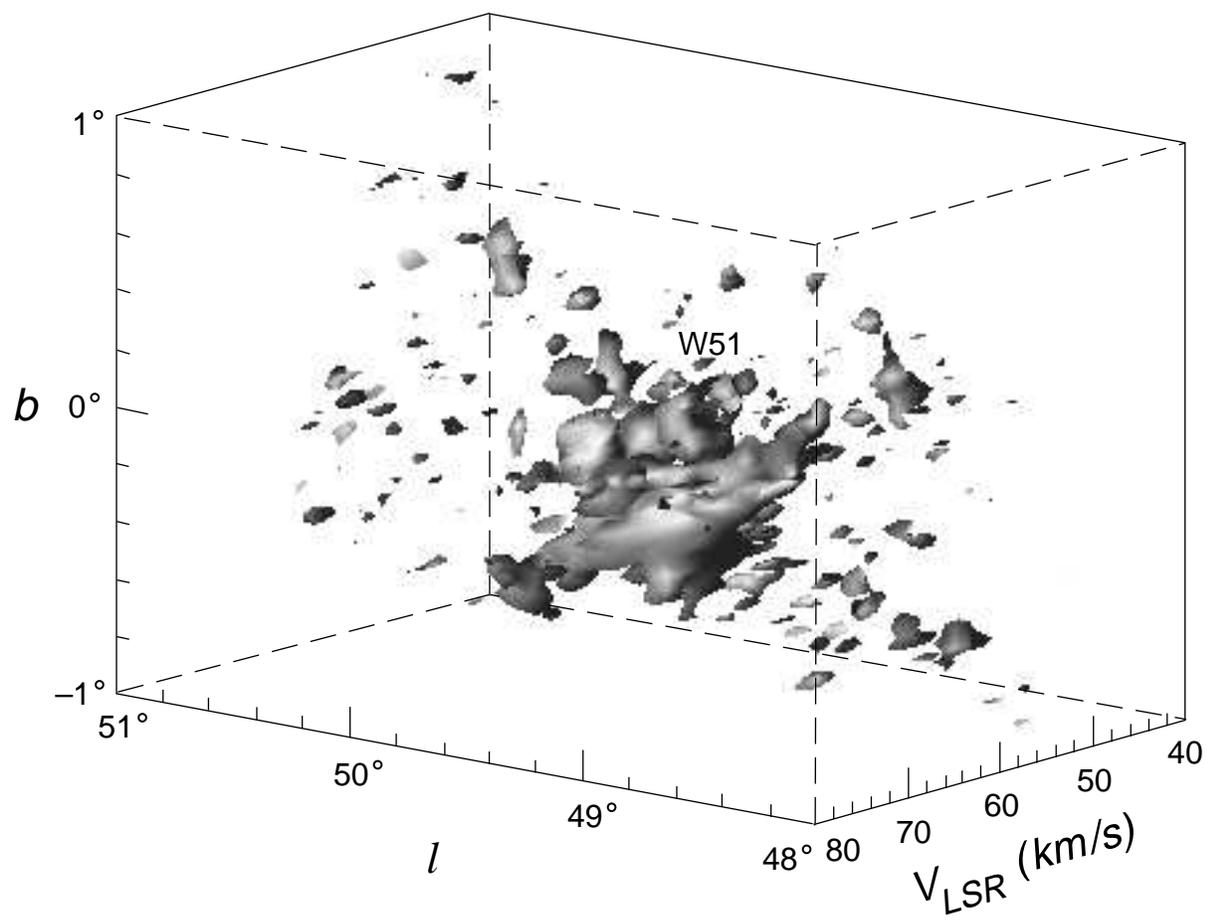
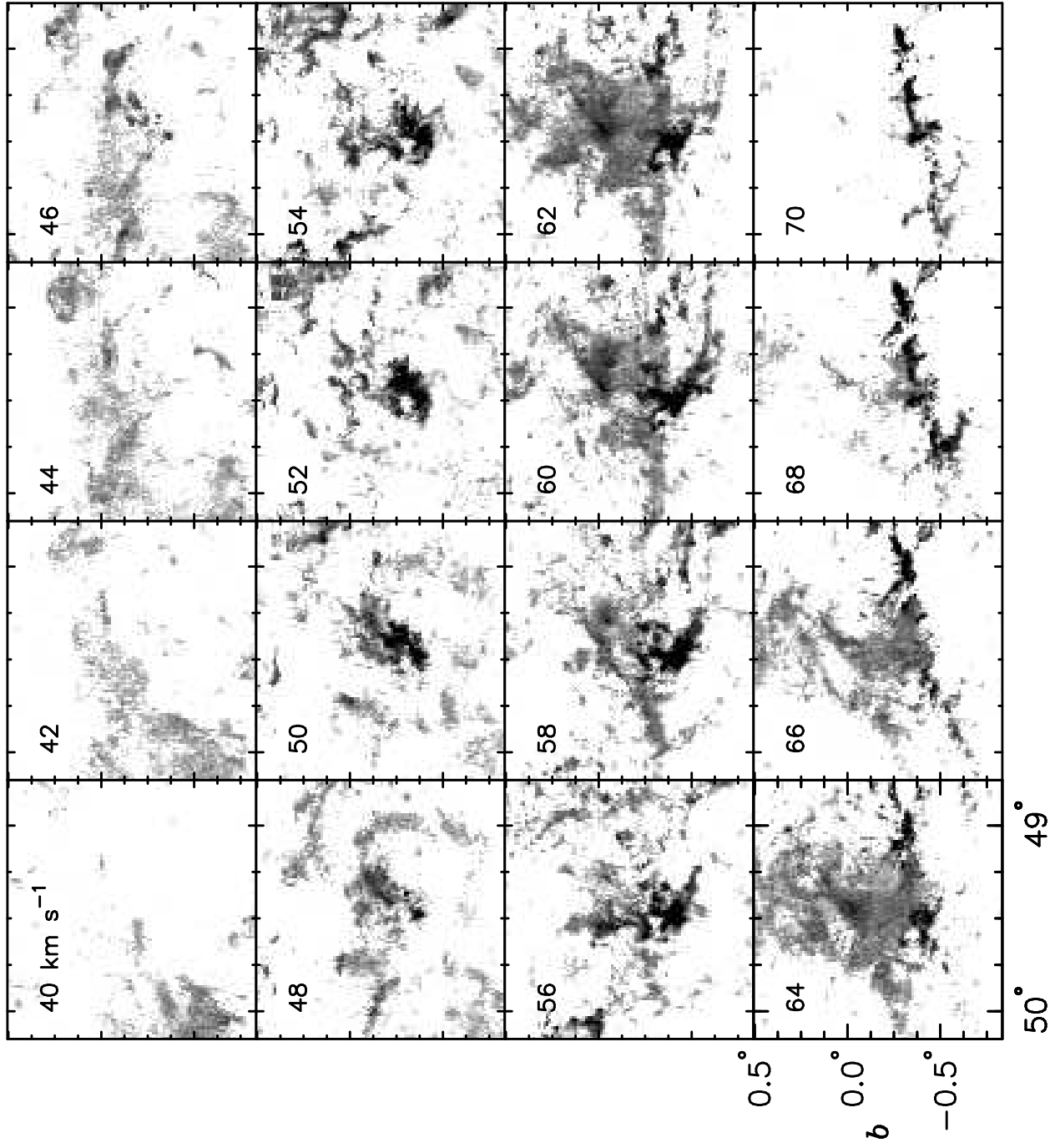
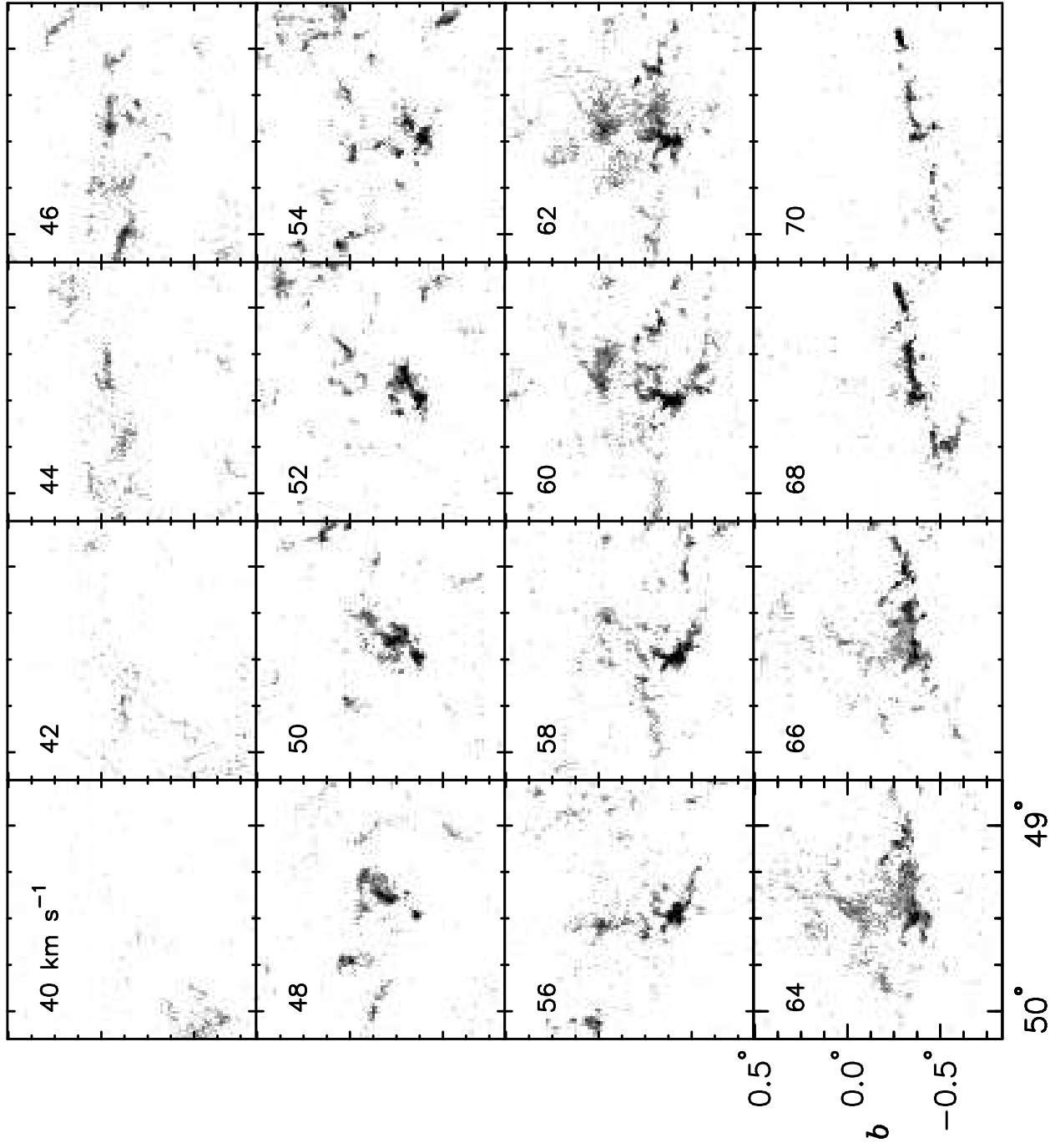
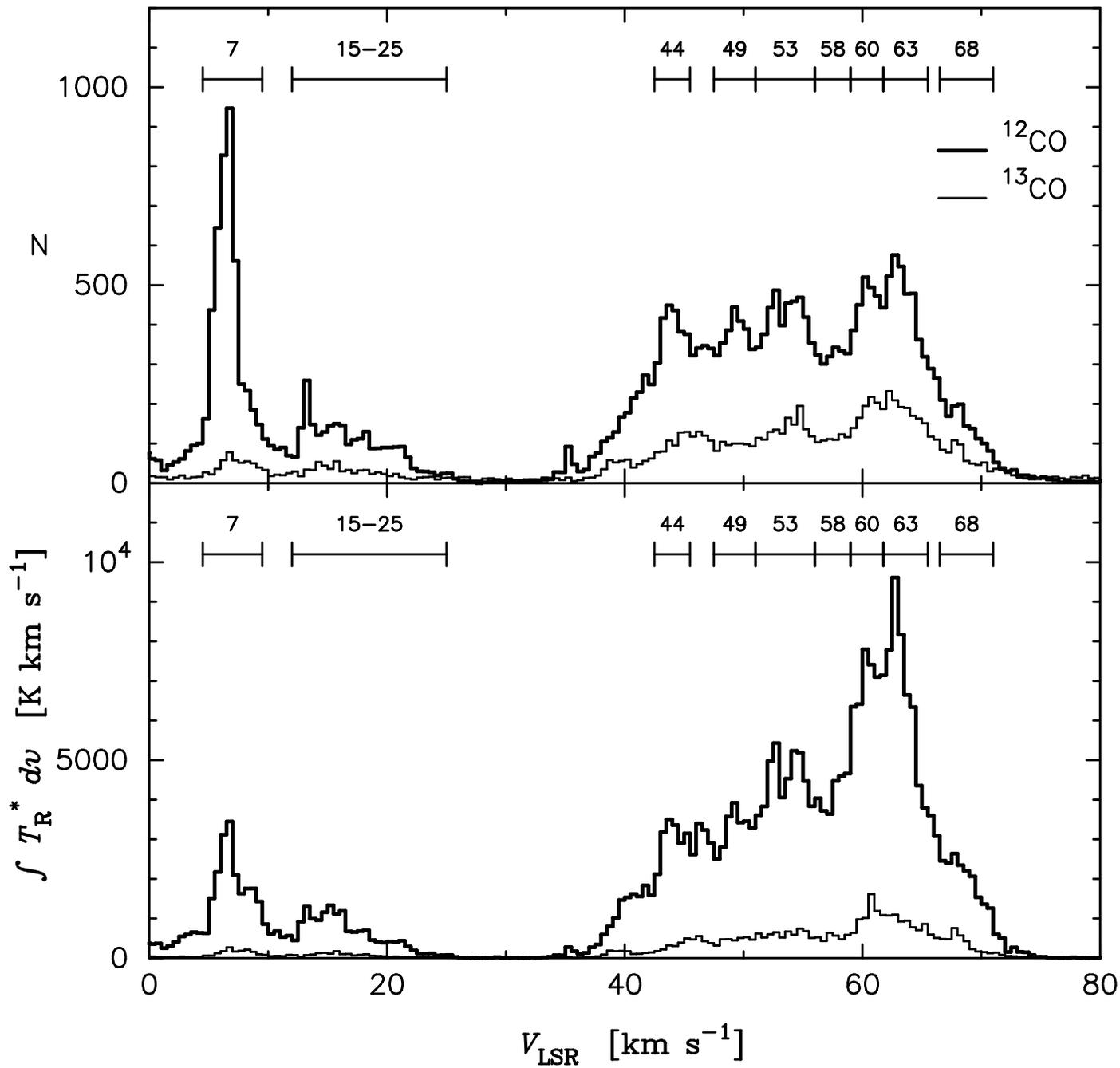
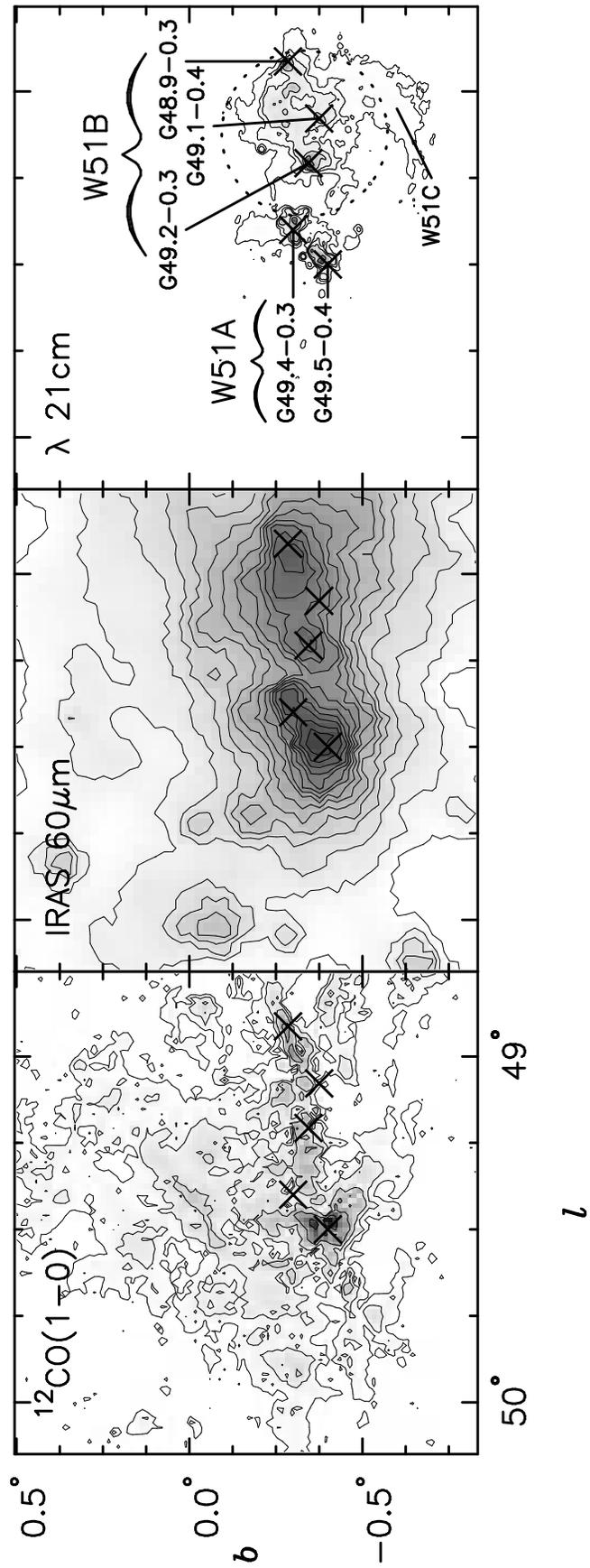
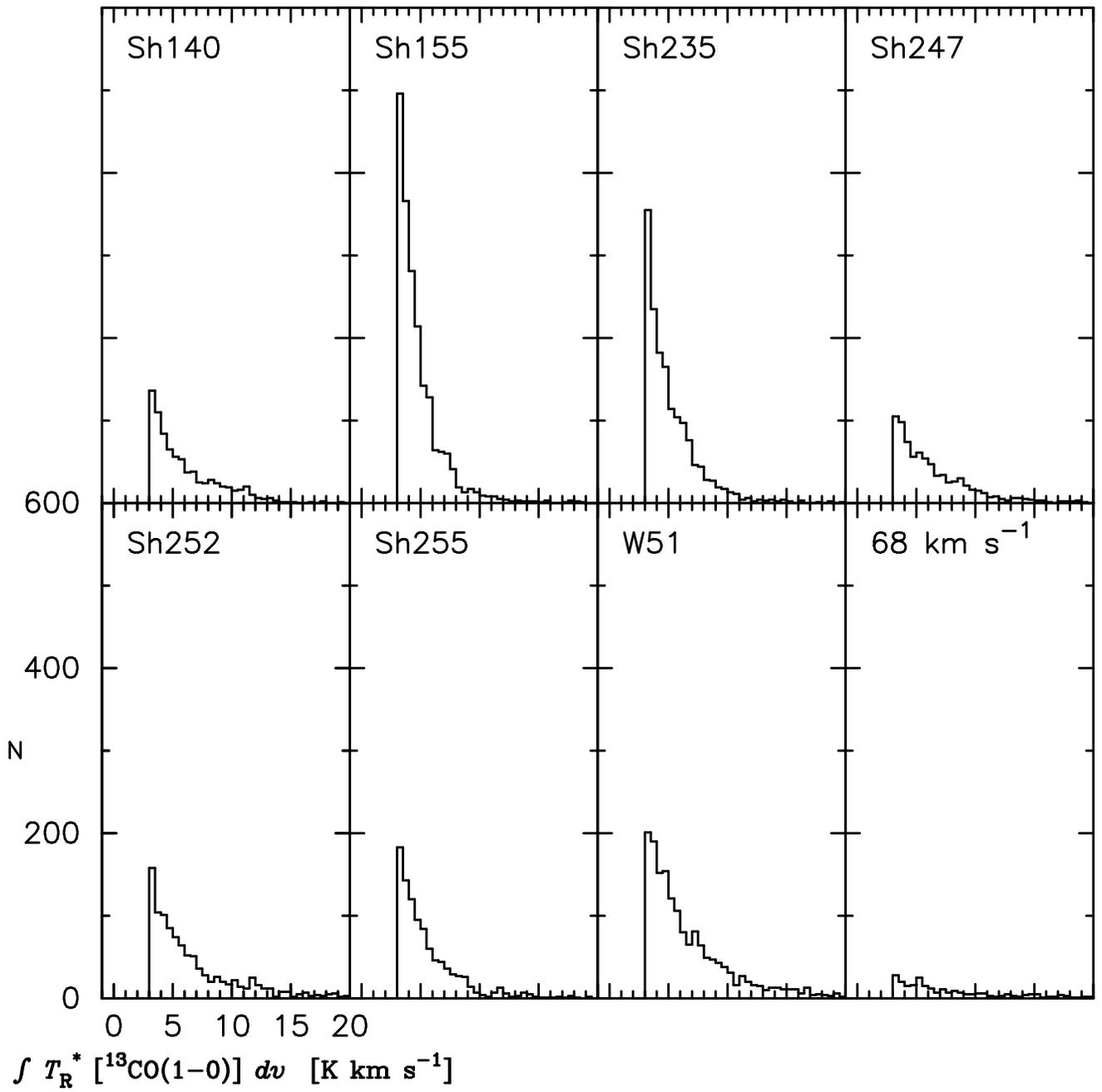
\end{document}